\documentclass[10pt,preprint]{emulateapj}

\slugcomment{}

\shorttitle{GRB Plateau: Hint for Solidification of Quark Matter?}
\shortauthors{Dai S., Xu R. X. and Li L. X.}

\begin{document}

\title{The Plateau of Gamma-ray Burst:
Hint for the Solidification of Quark Matter?}

\author{Shi Dai\altaffilmark{1}, Lixin Li\altaffilmark{2}, Renxin Xu\altaffilmark{1}}
\email{\{daishi, lxl, r.x.xu\}@pku.edu.cn\\
$^1$School of Physics and State Key Laboratory of Nuclear Physics
and Technology, Peking University, Beijing 100871,
China;\\
$^2$Kavli Institute for Astronomy and Astrophysics, Peking
University, Beijing 100871, China.}

\begin{abstract}
The origin of the shallow decay segment in gamma-ray burst's (GRB)
early light curves remains a mystery, especially those cases with a
long-lived plateau followed by an abrupt falloff. In this paper, we
propose a mechanism to understand the origin of the abrupt falloff
after plateau by considering solidification of newborn quark stars
with latent heat released as energy injection to GRB afterglow.
We estimate the total latent heat released during the phase transition
of quark stars from liquid to solid states, to be order of $\sim
10^{51}$ergs, which is comparable to the emission energy in the
shallow decay segment. We also estimate the time scale of radiating
the latent heat through thermal photon emission, and find that the
time scale agrees with observations.
Based on our estimation, we analyze the process of energy injection
to GRB afterglow. We show that the steady latent heat of quark star
phase transition would continuously inject into GRB afterglow in a
form similar to that of a Poynting-flux-dominated outflow and naturally
produce the shallow decay phase and the abrupt falloff after plateau.
We conclude that the latent heat of quark star phase transition
would be an important contribution to the shallow decay radiation in
GRB afterglow, and would explain the general features of GRB light
curves (including the plateau), if pulsar-like stars are really
(solid) quark stars.

\end{abstract}


\section{Introduction}
NASA's broadband (gamma-ray, X-ray, UV \& optical) Swift satellite's \citep{Gehrels2004}
successful launch and operation opens a brand new era in the observation
of the gamma-ray burst (GRB) phenomenon and has revolutionized our
understanding of GRBs in many aspects \citep{Meszaros2006, Zhang2007}.
The prompt slewing capability of Swift allows us to swiftly catch
the very early signals following the GRB prompt emission and the
precise localizations make it possible for ground-based follow up
observations of most bursts. A large number of well-sampled X-ray
light curves from tens of seconds to days past the GRB triggers
\citep{Burrows2004, Liang2007} have been accumulated through Swift's
observation, which provide us a good chance to investigate early
afterglow and study GRB systematically.

A canonical light curve of X-ray afterglows as revealed by Swift
\citep{Zhang2006, Nousek2006, O¡¯Brien2006} is composed of five
parts: prompt gamma-ray phase with a tail, shallow decay phase,
normal decay phase, jetlike decay phase, and erratic X-ray flares.
The physical origins of these segments have been widely discussed in
the literature \citep{Zhang2006, Nousek2006, Dai2006, Liang2007}. In this
paper, we mainly focus on the shallow decay phase and the subsequent
phase. On the one hand, the physical origin of the shallow decay
phase is still a mystery. Different models, such as the energy
injection models \citep{Zhang2006, Nousek2006, Panaitescu2006b}, a
combination of the GRB tail with delayed onset of the afterglow
emission \citep{Kobayashi2007}, off-beam jets \citep{Toma2006,
Eichler2006}, precursor activity \citep{Ioka2006}, two-component
jets \citep{Granot2006, Jin2007}, two-component emission model \citep{Yamazaki2009},
varying microphysical parameters \citep{Ioka2006, Panaitescu2006a, Fan2006, Granot2006}
and so on, are hard to differentiate among each other from the X-ray
observations \citep{Zhang2007}. On the other hand, some puzzling
facts related to the shallow decay phase are revealed by Swift's
observations, which can not be explained by current models. Two
interesting facts are that the break between the shallow and the
normal decay segments in the X-ray light curve for some GRBs is
chromatic \citep{Panaitescu2006a, Fan2006} and that the light curve
of GRBs like GRB 070110 shows a long-lived plateau followed by an
abrupt falloff (the decay slope is about $-9$, with time zero at the
trigger) \citep{Liang2007}. The former fact suggests that the
optical and X-ray emission in the shallow decay phase may not be the
same component, and the later fact indicates an internal origin of
the X-ray plateaus and continuous operation of a long-term central
engine. Systematically analysis of the Swift X-Ray Telescope data
suggests that the physical origin of the shallow decay phase is
diverse \citep{Liang2007} and we expect to get more information from
it including the physics of dense material and possible gravitational
wave \citep{Alessandra2009}.

We note that GRB central engine may relate to the physics of cold
matter at supra-nuclear density, which is now one of the daunting
challenges in particle physics. Cold quark matter is conjectured to
be in a solid state at realistic baryon densities of compact stars
\citep[e.g.,][for a review]{Xu2009}, we are then considering latent
heat of quark star phase transition from liquid to solid as energy
injection to GRB afterglow in order to understand the feature of
plateau followed by an abrupt falloff in some GRB's light curve.
Quark stars, as possible nature of pulsars, are likely to form in
GRB, no matter in the process of high-mass star collapses or merge
of binary neutron stars. In the solid quark star model, which is
successful to understand a variety of pulsars' observational
features \citep{Xu2003}, it is likely that as the temperature of
quark star drops after GRB, a phase transition from liquid to solid
would happen \citep{XuLiang2009}. Since the quark star in phase
transition would emit energy with constant temperature and solid
quark star would cool very fast due to its low heat capacity
\citep{Yu2009}, the latent heat of this quark star phase transition
not only provides a long-term steady central engine, but also
naturally shows a abrupt cutoff when the phase transition ends. In
this paper, we show that not only the energy released during the
phase transition and the time scale of radiating latent heat agree
with observations, but the process of energy injection to GRB
afterglow is also reasonable. Thus latent heat of quark star phase
transition as energy injection to GRB afterglow would be an
prospective model to understand the physical origin of the shallow
decay phase of GRB light curve, especially for the feature of abrupt
falloff after plateau.

In Section 2, we estimate the latent heat of quark star phase
transition in solid quark star model. Section 3 discusses the process
of energy injection. We conclude the results in Section 4.

\section{Latent heat of quark star phase transition}
To estimate the latent heat of quark star phase transition from
liquid to solid, we need to know the state of cold quark matter
and the interaction between quarks. However, due to the
non-perturbative effect of the strong interaction between quarks
at low energy scale and the many-body problem of vast assemblies
of interacting particles, we can not describe the state of cold
quark matter from first principle up to now. Yet, it is
phenomenologically conjectured that astrophysical cold quark matter
could be in a solid state, and a variety of observational features,
which may challenge us in the hadron star model, could be naturally
understood in the solid quark star model \citep{Xu2003}. Recent
results of relativistic heavy ion collision experiments also show
that the interaction between quarks is very strong in hot quark-gluon
plasma \citep{Shuryak2009}, then it is reasonable to conjecture that
the interaction between quarks should be stronger in cold quark matter.
The strong interaction may then make quarks grouped in clusters, and
if the residual interaction between quark clusters is stronger than
their kinetic energy, each quark cluster could be trapped in the
potential well and cold quark matter will be in a solid state \citep{Xu2009}.

Considering that a single quark cluster inside a quark star is
assumed to be colorless, just like each molecule in a bulk of inert
gas is electric neutral, Lai and Xu in their recent work \citep{Lai2009}
used Lennard-Jones potential to describe the interaction between
quark clusters in quark stars and got the equation of state of
quark stars. The interaction is expressed as
\begin{equation}u(r)=4U_0[(\frac{r_0}{r})^{12}-(\frac{r_0}{r})^6],\end{equation}
where $U_0$ is the depth of the potential and $r_0$ can be
considered as the range of interaction.

Based on this solid quark star model, we can then estimate
the latent heat of quark star phase transition from liquid to
solid. Rather than performing difficult molecular dynamics
simulations of crystallization, we would prefer estimating
the latent heat in order of magnitude by analogy with inert
gas and common substances since the quark clusters are non-relativistic
and the interaction is similar to common substances.
In Table$1$ below, we list the melting heat, heat of vaporization,
potential and ratio of melting heat to potential of inert
gas and some common substances. The melting heat corresponds
to the latent heat, and for substances with known heat of
sublimation, we equal the potential to heat of sublimation,
otherwise we equal the potential to the sum of melting heat and
heat of vaporization.

From the data, we can see that for most substances the ratio of
melting heat to potential is between $0.1$ to $0.01$. Considering
that the interaction between quark clusters is similar to inert
gas and is relatively strong, we choose the ratio of potential to
melting heat to be $f\approx0.1\sim0.01$ for estimation. Then
based on the solid quark star model proposed by Lai, choosing
$U_0=100$MeV \citep{Lai2009}, we can estimate the energy released
by each quark cluster in the liquid to solid phase transition as
\begin{equation}
E_{\rm cluster}\sim fU_{0}\approx 1\sim10 \rm{MeV}.
\end{equation}
For quark star of one solar mass, $M_{\odot}\approx 2\times10^{33}$g,
the number of baryon is $n=10^{57}$, then the total energy released
during the phase transition can be estimated as
\begin{equation}
E=E_{\rm cluster}n\approx 10^{51}\sim10^{52}\rm{ergs}.
\end{equation}
This order of magnitude agrees with the typical energy released
in the shallow decay phase of GRB, that is to say, the latent
heat of quark star phase transition from liquid to solid is
sufficient to produce the plateau.

On the other hand, according to the Lindemann law that a solid
melts when the root-mean-square amplitude of atomic vibrations
exceeds a certain fraction of the equilibrium nearest neighbor
distance, we can estimate the temperature of quark star phase
transition and further more the time scale of radiating latent
heat. In Mohazzabi and Behroozi's work in 1987, they obtained
the expression of the ratio of the root-mean-square amplitude
of atomic vibrations to the equilibrium nearest neighbor
distance for inert gas, and found that the Lindemann law being
well consistent with experiments \citep{Mohazzabi1987}. As for
our estimation, we can consult the ratio of potential to heat,
$\Gamma=U_{0}/kT$, for common substances, and then get the
temperature of quark star phase transition by analogy. For
one-component plasma, $\Gamma\approx175$ \citep{DeWitt2001},
for multi-component plasma, $\Gamma\approx233$ \citep{Horowitz2007}.
So choosing $\Gamma\approx200$, $U_0=100$MeV, the temperature
of quark star phase transition is around $0.5$MeV.

During the liquid to solid phase transition, the temperature of
quark star would remain constant, and the latent heat would
be released through thermal emission. Then the time scale of radiation
can be estimated as
\begin{equation}
t=\frac{E}{\sigma T^{4} 4\pi R^{2}},
\end{equation}
where, $E=10^{51}$ ergs, $\sigma$ is the Stefan-Boltzman constant,
$R=10$km is the radius of quark star, and $T\approx0.5$MeV, we
find that the time
scale of radiation is $t\approx1000$s, which agrees with
observations of GRB afterglow plateau.

After solidification, since the heat capacity of solid quark stars
is very low \citep{Yu2009}, the central quark star would cool rapidly.
The residual inner energy would be released almost immediately, thus
an abrupt cutoff of energy injection to the afterglow would appear, and
naturally result in an abrupt falloff after the shallow decay phase.
In Fig$1$, a schematic cooling curve of quark star is presented, which
consists of three stages. The first stage could be an initial fast cooling
stage due to the emission of neutrinos and photons at the very beginning
of a quark star, when its initial temperature $T_{0}$ could be much higher
than $10$MeV. When the temperature of quark star drops to the melting
point $T_{p}$, it could come to the second stage, the liquid to solid
phase transition which last from $t_{i}$ to $t_{f}$. At this stage,
the temperature of quark star would remain constant and the latent heat
would be released steadily which could provide a long-lived steady central
engine. After phase transition, the born solid quark star would rapidly
release its residual inner energy due to its low heat capacity, thus a
steep falloff appears in the cooling curve corresponding to the abrupt
falloff after the shallow decay segment.

\section{The standard fireball revisited?}

In the standard GRB fireball model, it is assumed that a large amount
of energy is instantly released through some explosion process. The energy
drives some material to expand with a ultra-relativistic speed, which requires
that the {\em fireball} outflow is low-baryon-loaded so that the total rest
mass of the fireball is sufficiently small. The standard fireball is assumed
to be highly non-uniform and in the extreme case is composed of many distinct
shells. Collision of different shells produces internal shock waves, which
are thought to be responsible for the observed prompt emission. Collision of
shells with the surrounding intermediate stellar material produces external
shock waves, which are thought to be responsible for the observed GRB afterglow
emission. While for the afterglow flares, they are usually thought to be
produced by later injection of energy through internal shock.

After having estimated the energy released by quark star phase
transition and the time scale of radiation, it is necessary for
us to qualitatively discuss the process of energy injection to
GRB afterglow. Generally speaking, the energy injection to
afterglow could consist of some kinetic-energy (i.e., baryons)
dominated shells or a Poynting-flux-dominated wind \citep{Usov1994, Meszaros1997}.
In case of afterglow, we always consider an impulsive
shell that is already heated during the shell-ISM interaction
and that is collecting material from the ISM, and in the
meantime also receives a large enough injection energy
from a continuous Poynting-flux-dominated wind or a kinetic-energy
dominated shell. As discussed in literatures \citep{Zhang2001, Zhang2002},
for Poynting-flux-dominated wind case in which pure energy with
negligible baryon loading is injected to fireball, no
reverse shock is expected, and the injection signature
is produced only by the forward shock emission. For
kinetic-energy-dominated matter shells case, depending on
whether the collision between the injected and the impulsive
shell is mild or violent, the injection process is quite
different. If the relative velocity between the colliding
shells does not exceed a critical value defined by their energy
ratio, the collision is mild, and the injection may be analogous
to the Poynting-flux injection case. Otherwise, the injection
is violent, and an additional pair of strong shocks will form at
the discontinuity between two colliding shells which will greatly
influence the injection signature.

As for our energy injection model, the latent heat released during
quark star phase transition would produce a radiation-dominated
fireball, whose luminosity is about $L\approx10^{48}$ergs/s
according to our estimation. The optical depth can then be
estimated as
\[
\tau_{\gamma\gamma}=\frac{f_{p}\sigma_{T}FD^{2}}{R^{2}m_{e}c^{2}}
\]
\begin{equation}
=10^{13}f_{p}(\frac{D}{3000\rm{Mpc}})^{2}(\frac{F}{10^{-11}\rm{ergs/cm^{2}}})(\frac{R}{10\rm{km}})^{-2},
\end{equation}
where, $R=10$km is the radius of quark star. The optical depth is
very large, so the energy flux from central quark star would form a
pure radiation fireball consist of radiation and electron-positron pairs,
which is similar to the fireball of prompt emission but continuous instead.
We expect that this pure radiation fireball would expand and finally
inject energy to the afterglow in form of Poynting flux. If we consider
possible baryons in the space the fireball swept, we can estimate the velocity
the matter shell could reach assuming that all the energy are converted
to the kinetic energy of the baryons. Since in this late injection phase
the baryon loading could be in principle much lower, we assume that the
density of baryons are $10\%$ of the original density in space.
Then we can estimate as
\begin{equation}
\gamma=E/Mc^{2}.
\end{equation}
For $E\approx10^{51}$ergs, we get $\gamma\approx100$. According
to Zhang and M\'{e}sz\'{a}ros' work in 2002 \citep{Zhang2002},
for such kinetic-energy-dominated matter shells with
$\gamma\approx100$ and $L\approx10^{48}$ergs/s, the collision
between the injected and the impulsive shell is mild, and the
injection may be analogous to the Poynting-flux injection case.

\section{Conclusions}

A possible physical origin of the shallow decay phase of GRB light
curve is proposed in the solid quark star model. We suggest that
quark stars may be born in GRB, and as the temperature of quark star
drops, a phase transition from liquid to solid may occur in the
quark star. The latent heat of phase transition would provide a
long-lived steady energy injection to GRB afterglow, since the
temperature of central star would remain constant during phase
transition. When phase transition ends, an abrupt falloff after the
plateau in the light curve would naturally appear which is hard to
understand by other central engine models.

We estimate the latent heat of quark star phase transition from
liquid to solid based on the solid quark star model whose
interaction between quark clusters is described by Lennard-Jones
potential. The energy of $\sim 10^{51}$ ergs and the radiation time
scale of $\sim 10^3$ seconds agree with observations of the shallow
decay phase. We also qualitatively discuss the process of energy
injection to afterglow, and show that the energy injection of phase
transition would be in a form similar to that of the
Poynting-flux-dominated outflow. Both the estimation and the
injection process suggest that the idea of considering latent heat
of quark star phase transition as energy injection to GRB afterglow
is rational.

\acknowledgments We would like to acknowledge useful discussions at
our pulsar group of PKU. This work is supported by the National
Natural Science Foundation of China (10973002, 10935001), the
National Basic Research Program of China (2009CB824800), and the
National Fund for Fostering Talents of Basic Science (J0630311).

\clearpage

\begin{figure*}
\caption{A schematic cooling behavior of a new-born quark star.
Three stages are shown: an initial cooling stage due to the emission
of neutrinos and photons at the very beginning of a quark star with
initial temperature $T_{0}$, a liquid to solid phase transition
stage from time $t_{i}$ to $t_{f}$ with constant temperature
$T_{p}$, and, after solidification, a fast cooling stage because of
solid quark star's low heat capacity. We focus on the duration from
$t_i$ to $t_f$ in this paper.} \centering
\includegraphics[height=7cm]{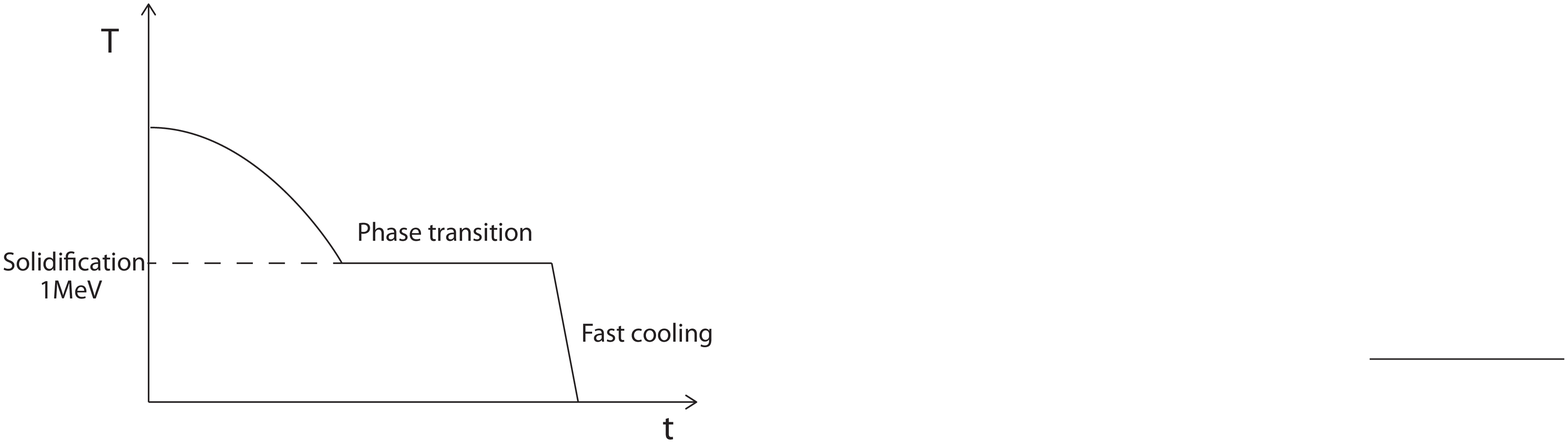}
\end{figure*}

\begin{table*}
\caption{The melting heat, heat of vaporization,
potential and ratio of melting heat to potential of inert
gas and some common substances \citep{Dean1999}.}
\centerline{\begin{tabular}{|c|c|c|c|c|c|}\hline
 &  melting heat kcal/mol  & heat of vaporization kcal/mol   & potential kcal/mol  &   melting heat/potential  \\
\hline
He & 0.0033 & 0.0194 & 2.2944   & 0.0014 \\
\hline
Ne & 0.0801 & 0.422  & 9.1776   & 0.0087 \\
\hline
Xe & 0.5495 & 3.02   & 12.6192  & 0.0436  \\
\hline
Rn & 0.69   & 4.01   & 19.5024  & 0.0354  \\
\hline
   &        &        &          &        \\
\hline
Al & 2.56   & 69.5   & 78       & 0.0328  \\
\hline
Cs & 0.499  & 16.198 & 18.3     & 0.0272  \\
\hline
Cu & 3.17   & 72.74  & 81       & 0.0391  \\
\hline
Fe & 3.63   & 83.68  & 99.5     & 0.0365  \\
\hline
Hg & 0.5486 & 14.13  & 14.65    & 0.0375   \\
\hline
Na & 0.622  & 23.285 & 25.75    & 0.0242   \\
\hline
Si & 12     & 85.8   & 107.7    & 0.1114  \\
\hline
C  & 25     &        & 171.29   & 0.1460   \\
\hline
CO & 0.2    & 1.444  & 1.644    & 0.1217   \\
\hline
$\rm{CO}_{2}$  & 1.99      &        & 6.03   & 0.3300  \\
\hline
$\rm{H}_{2}O$  & 1.436      &  9.717    & 11.153   & 0.1287  \\
\hline
$\rm{H}_{2}\rm{O}_{2}$  & 2.987   &  10.53   & 12.34   & 0.2421   \\
\hline
$\rm{CaCl}_{2}$  & 6.8      & 56.2       & 77.5   & 0.0877  \\
\hline
\end{tabular}}
\end{table*}


\begin{thebibliography}{}

\bibitem[Alessandra \& M\'{e}sz\'{a}ros(2009)]{Alessandra2009}
Alessandra, C., \& M\'{e}sz\'{a}ros, P. 2009, ApJ, 702, 1171

\bibitem[Burrows et al.(2004)]{Burrows2004}
Burrows, D. N., et al. 2004, Proc. SPIE, 5165, 201

\bibitem[Dai et al.(2006)]{Dai2006}
Dai, Z. G., Wang, X. Y., Wu, X. F., Zhang, B. 2006, Science, 311, 1127

\bibitem[Dean(1999)]{Dean1999}
Dean, J. A. 1999, Lange's Chemistry Handbook (Version 15th, in Chinese), Table 6.2.

\bibitem[DeWitt et al.(2001)]{DeWitt2001}
DeWitt, H., et al. 2001, Contrib Plasma Phys, 41, 251

\bibitem[Eichler \& Granot(2006)]{Eichler2006}
Eichler, D., \& Granot, J. 2006, ApJ, 641, L5

\bibitem[Fan \& Piran(2006)]{Fan2006}
Fan, Y., \& Piran, T. 2006, MNRAS, 369, 197

\bibitem[Gehrels et al.(2004)]{Gehrels2004}
Gehrels, N. et al. 2004, ApJ, 611, 1005.

\bibitem[Granot et al.(2006)]{Granot2006}
Granot, J., K\"{o}nigl, A., \& Piran, T. 2006, MNRAS, 370, 1946

\bibitem[Horowitz et al.(2007)]{Horowitz2007}
Horowitz, C. J., et al. 2007, Physical Review E, 75, 066101

\bibitem[Ioka et al.(2006)]{Ioka2006}
Ioka, K., Toma, K., Yamazaki, R., \& Nakamura, T. 2006, A\&A, 458, 7

\bibitem[Jin et al.(2007)]{Jin2007}
Jin, Z.-P., Yan, T., Fan, Y.-Z., \& Wei, D.-M. 2007, ApJ, 656, L57

\bibitem[Kobayashi \& Zhang(2007)]{Kobayashi2007}
Kobayashi, S., \& Zhang, B. 2007, ApJ, 655, 973

\bibitem[Lai \& Xu(2009)]{Lai2009}
Lai, X. Y., \& Xu, R. X. 2009, MNRAS, 398, 31

\bibitem[Liang et al.(2007)]{Liang2007}
Liang, E. W., Zhang, B. B., Zhang, B. 2007, ApJ, 670,565

\bibitem[Lindemann(1910)]{Lindemann1910}
Lindemann, F. A. 1910, Physik Z., 11, 609

\bibitem[M\'{e}sz\'{a}ros(2006)]{Meszaros2006}
M\'{e}sz\'{a}ros, P. 2006, Rep. Prog. Phys., 69, 2259

\bibitem[M\'{e}sz\'{a}ros \& Rees(1997)]{Meszaros1997}
M\'{e}sz\'{a}ros, P., \& Rees,M. J. 1997, ApJ, 482, 29

\bibitem[Mohazzabi \& Behroozi(1987)]{Mohazzabi1987}
Mohazzabi, P. \& Behroozi, F. 1987, Journal of Materials Science Letters, 6, 404

\bibitem[Nousek et al.(2006)]{Nousek2006}
Nousek, J. A., et al. 2006, ApJ, 642, 389

\bibitem[O¡¯Brien et al.(2006)]{O¡¯Brien2006}
O¡¯Brien, P. T., et al. 2006, ApJ, 647, 1213

\bibitem[Panaitescu et al.(2006a)]{Panaitescu2006a}
Panaitescu, A., M\'{e}sz\'{a}ros, P., Burrows, D., Nousek, J., Gehrels, N., O¡¯Brien, P.,
\& Willingale, R. 2006a, MNRAS, 369, 2059

\bibitem[Panaitescu et al.(2006b)]{Panaitescu2006b}
Panaitescu, A., M\'{e}sz\'{a}ros, P., Gehrels, N., Burrows, D., \& Nousek, J. 2006b, MNRAS, 366, 1357

\bibitem[Shuryak(2009)]{Shuryak2009}
Shuryak E. V., 2009, Prog. Part Nucl. Phys., 62, 48

\bibitem[Toma et al.(2006)]{Toma2006}
Toma, K., Ioka, K., Yamazaki, R., \& Nakamura, T. 2006, ApJ, 640, 139

\bibitem[Usov(1994)]{Usov1994}
Usov, V. V. 1994, MNRAS, 267, 1035

\bibitem[Xu(2003)]{Xu2003}
Xu, R. X. Solid quark stars. 2003, ApJ, 596, 59

\bibitem[Xu(2010)]{Xu2009}
Xu R. X., 2010, IJMP D (arXiv:1002.4469, in press)

\bibitem[Xu \& Liang(2009)]{XuLiang2009}
Xu, R. X., \& Liang, E. W. 2009, Sci China Ser G-Phys Mech Astron, 52, 315

\bibitem[Yamazaki(2009)]{Yamazaki2009}
Yamazaki, R. 2009, ApJ, 690, 118

\bibitem[Yu \& Xu(2009)]{Yu2009}
Yu, M., \& Xu, R. X. 2009, arXiv:0905.3818

\bibitem[Zhang(2007)]{Zhang2007}
Zhang, B. 2007, Chinese J. Astron. Astrophys., 7, 1

\bibitem[Zhang \& M\'{e}sz\'{a}ros(2001)]{Zhang2001}
Zhang, B., \& M\'{e}sz\'{a}ros, P. 2001, ApJ, 552, 35

\bibitem[Zhang \& M\'{e}sz\'{a}ros(2002)]{Zhang2002}
Zhang, B., \& M\'{e}sz\'{a}ros, P. 2002, ApJ, 566, 712

\bibitem[Zhang et al.(2006)]{Zhang2006}
Zhang, B., Fan, Y.-Z., Dyks, J., Kobayashi, S., M\'{e}sz\'{a}ros, P., Burrows, D. N.,
Nousek, J. A., \& Gehrels, N. 2006, ApJ, 642, 354

\end{thebibliography}
\end{document}